\documentclass{aastex63}

\usepackage{amsmath,amssymb}
\usepackage{relsize}
\usepackage{wasysym}
\usepackage{xspace}
\usepackage[normalem]{ulem}
\usepackage{upgreek}
\usepackage{nicefrac}
\usepackage{pgffor}
\makeatletter
\def\mitya{\@ifnextchar[{\@mwith}{\@mwithout}}

\def\@mwith[#1]#2{
  {\color{blue}  #2}\marginpar{$\color{blue}\cal M$}\footnote{Comment: #1}
}

\def\@mwithout#1{
  {\color{blue}  #1}\marginpar{$\color{blue}\cal M$}
}

\def\myd{\mathrm{d}}
\def\myd{{\rm d}}
\def\dif{\@ifnextchar[{\@with}{\@without}}

\def\@with[#1]#2{
  \ensuremath{
    \mathchoice
    {\frac{\foreach \x in {#2}{\myd\x\,}}{\foreach \x in {#1}{\myd\x\,}}}%
    {{\foreach \x in {#2}{\myd\x\,}}/{\foreach \x in {#1}{\myd\x\,}}}%
    {{\foreach \x in {#2}{\myd\x\,}}/{\foreach \x in {#1}{\myd\x\,}}}%
    {{\foreach \x in {#2}{\myd\x\,}}/{\foreach \x in {#1}{\myd\x\,}}}
  }
}

\def\@without#1{
  \ensuremath{%
    \ifx\hfuzz#1\hfuzz
    \myd
    \else
    \foreach \x in {#1}{\myd\x\,}
    \fi
    }
}
\makeatother

\newif\iflong
\longfalse

\newcommand{\be}{\begin{equation}}
\newcommand{\ee}{\end{equation}}
\newcommand{\ba}{\begin{eqnarray}}
\newcommand{\ea}{\end{eqnarray}}

\newcommand{\sigmat}{\ensuremath{\sigma_\textsc{t}}}

\newcommand{\syn}{\ensuremath{_\textsc{syn}}}

\newcommand{\acc}{\ensuremath{_\textsc{acc}}}
\newcommand{\eff}{\ensuremath{_\textsc{eff}}}

\newcommand{\ma}{\ensuremath{^\textsc{max}}}
\newcommand{\mi}{\ensuremath{^\textsc{min}}}
\newcommand{\me}{\ensuremath{m_{\rm e}}}
\newcommand{\Ee}{\ensuremath{E_{\rm e}}}

\newcommand{\dias}{{Dublin Institute for Advanced Studies, School of Cosmic Physics, 31 Fitzwilliam Place, Dublin 2, Ireland}}

\newcommand{\mpik}{{Max-Planck-Institut f\"ur Kernphysik, Saupfercheckweg 1, 69117 Heidelberg, Germany}}
\newcommand{\rikkyo}{{Department of Physics, Rikkyo University, Nishi-Ikebukuro 3-34-1, Toshima-ku, Tokyo 171-8501, Japan}}

\newcommand{\desy}{{DESY, D-15738 Zeuthen, Germany}}

\received{January 15, 2021}
\revised{January 10, 2019}
\accepted{\today}
\submitjournal{ApJ}

\shorttitle{Extension of the synchrotron radiation of electrons to very high energies in clumpy environments}
\shortauthors{Khangulyan et al.}

\begin{document}

\title{Extension of the synchrotron radiation of electrons to very high energies in clumpy environments}

\correspondingauthor{Dmitry Khangulyan}
\email{d.khangulyan@rikkyo.ac.jp}

\author[0000-0002-7576-7869]{Dmitry Khangulyan}
\affiliation{\rikkyo}

\author{Felix Aharonian}%
\affiliation{%
 \dias
}%
\affiliation{%
 \mpik
}%


\author{Carlo Romoli}
\affiliation{
  \mpik
}%

\author{Andrew Taylor}
\affiliation{
  \desy
}%



\begin{abstract}
  The synchrotron cooling of relativistic electrons is one of the most effective radiation mechanisms in astrophysics. It not only accompanies the process of particle acceleration but also has feedback on the formation of the energy distribution of the parent electrons. The radiative cooling time of electrons decreases with energy as $t\syn \propto 1/E$; correspondingly the overall radiation efficiency increases with energy. On the other hand, this effect strictly limits the maximum energy of individual photons.  Even in the so-called extreme accelerators, where the acceleration proceeds at the highest possible rate, $t\acc^{-1} = eBc/E$, allowed in an ideal magnetohydrodynamic plasma, the synchrotron radiation cannot extend well beyond the characteristic energy determined by the electron mass and the fine-structure constant: $h \nu\ma \sim \me c^2/\alpha \sim 70 \rm\,MeV$.
  In this paper, we propose a model in which the formation of synchrotron radiation takes place in compact magnetic blobs located inside the particle accelerator and develop a formalism for calculations of synchrotron radiation emerging from such systems. We demonstrate that for certain combinations of parameters characterizing the accelerator and the magnetic blobs, the synchrotron radiation can extend beyond this limit by a several orders of magnitude. This scenario requires a weak magnetization of the particle accelerator, and an efficient conversion of gas internal energy into magnetic energy in sufficiently small blobs. The required size of the blobs is constrained by the magnetic mirroring effect, that can prevent particle penetration into the regions of strong magnetic field under certain conditions.


  
\end{abstract}



\section{Introduction} \label{sec:intro} Acceleration of non-thermal particles in various astrophysical environments
requires the presence of electric field.  If high-energy particles are accelerated in a region with a typical magnetic
field, \(B_0\), then the time, required for the particle to gain energy \(E\), is \(t\acc={\eta E}/({eB_0c})\). Here,
\(\eta\) is a dimensionless parameter, namely the ratio of the magnetic to the accelerating electric field,
\(\eta=B_0/{\cal E}_0\). If the acceleration proceeds in a magnetohydrodynamic (MHD) flow, then \({\cal E}_0<B_0\) and
the acceleration efficiency is constrained by \(\eta>1\). The acceleration is efficient only in the energy range where
the acceleration time is shorter than the cooling time. For the case where synchrotron losses dominate, the cooling time
is \(t\syn={6\pi \me^2c^3}/{(\sigmat B_0^2E)}\) (here \(\me\), \(c\), and \(\sigmat\) are electron mass, speed of light,
and Thomson cross-section, respectively). The electrons are subsequently accelerated up to a maximum energy
\(\Ee\ma=\sqrt{6\pi e\me^2c^4/(\sigmat \eta B_0)}\).  Correspondingly, the electrons which achieve this energy, produce
synchrotron photons of energy \(h \nu\ma \approx \me c^2/(\alpha\eta)\approx 70\eta^{-1}\rm\,MeV\) \citep[here
\(\alpha\) is the fine-structure constant]{1983MNRAS.205..593G}. {We note that this energy corresponds to the
  spectral maximum of synchrotron emission produced by an electron of energy \(\Ee\ma\) moving in a magnetic field
  \(B_0\) perpendicular to the velocity of the electron.  Various factors can shift the exact position of the synchrotron
  maximum (for example, particle velocity -- magnetic field pitch angle; turbulent dispersion of magnetic field
  strength; energy distribution of emitting particles; etc). However, the impact of these factors is expected to be
  small \cite[see, e.g.,][for the comparison of the spectra averaged over pitch angle and turbulent distribution of
  magnetic field strength]{2019ApJ...887..181D}.  A much more important factor is that this limit was obtained under the
  assumption of the perfect operation of the acceleration process, which is likely not achievable in real astrophysical
  accelerators. Thus, we adopt the above obtained energy as an energy that limits the extension of the synchrotron component (which is sometimes referred in
  literature as {\it synchrotron burnoff limit}, see, e.g., \citealt{1996ApJ...457..253D}). Finally, we note that the maxim of
  the spectral energy distribution (SED) appears at higher energies, namely for the same underlying assumptions at
  \(\approx 300\eta^{-1}\rm\,MeV\).
  
 There are several ways to
  alleviate this constraint on the cutoff energy of synchrotron component. Examples of a bypass are the inclusion of relativistic motion of the production site, acceleration
  with efficiencies exceeding the ideal MHD threshold (\(\eta<1\)), synchrotron emission of secondary electron positron pairs, synchrotron emission by particles heavier than electrons (muons and protons), and special regimes of emission (for example, in small-scale turbulent magnetic field). We discuss these approaches in
  Sec. \ref{sec:mhd_limit} in more detail. }
In this paper, we explore a new scenario for boosting the energy range of the synchrotron radiation utilising the
structure of the ambient magnetic field in the acceleration region. The principle behind our proposed scheme is the
separation of the region of the electron acceleration and the region(s) of effective synchrotron radiation. More
specifically, we consider a scenario in which the acceleration region contains a large number of compact magnetic field
condensations (blobs), whose collective volume filling factor is small. Provided that the
particle-blob collision time is long compared to the acceleration time, these blobs have only a minor impact on the
acceleration process. On the other hand, high-energy particles that eventually collide with these blob regions give rise
to a synchrotron radiation component which can extend well beyond 100~MeV. Throughout the discussion, we do not specify
the acceleration mechanism, as well as the nature of the compact magnetic condensations, but focus on the propagation
and radiation of the accelerated electrons colliding with the dense magnetic blobs. {A high degree of inhomogeneity of
  magnetic plasma is favored by numerical simulations of shocks, magnetic reconnection, and turbulence. The general consideration of the suggested
  scenario is therefore potentially relevant for a number of astrophysical systems.}

\section{Magnetobremsstrahlung feedback on particle spectrum}
\label{sec:mhd_limit}
The synchrotron burnoff turnover is caused by the feedback of synchrotron cooling on particle acceleration in ideal MHD
plasma and should be applied in the plasma co-moving reference frame. In the most simple setup, where the magnetic field
is homogeneous and perpendicular to the particle velocity, the cutoff should appear at
\(h \nu\ma \sim 70 {\cal D} m/(\me \eta)\rm\,MeV\), where \(m\) and \({\cal D}\) are mass of the emitting particle and
the Doppler boosting factor, respectively.  The Doppler boosting effect, which accounts for relativistic motion of the
production site, can be sufficient to explain, in a framework of electron synchrotron scenario, bright Crab GeV flares,
whose SED peaks at a few hundred MeV, \(h \nu\ma \lesssim400\rm\,MeV\). However
extension of the synchrotron component to the very high energy (VHE) regime requires an extremely high boosting factors
\({\cal D}\sim{\cal O}(10^3)\). Such fast outflows are not expected in astrophysical objects unless one considers
production of the emission in pulsar winds,  {whose bulk Lorentz factor can be very high, \({\cal D}\sim10^6\). These winds  originate in a very compact region, comparable in size to the
  pulsar magnetosphere, and almost unavoidably undergo strong adiabatic cooling due to their expansion. As a result, the
  temperature of the wind electrons (i.e., their energy in the co-moving frame) is expected to be very low, resulting in a
  strong suppression of the synchrotron emission. If some acceleration process, even with a modest efficiency, \(\eta<10^3\), operates in a pulsar wind, then the synchrotron emission produced
  there could be boosted into the VHE  regime.}

{Another obvious way to generate synchrotron emission in the VHE domain is to consider higher-mass particles. For
  example, protons, whose mass is \(\approx2000\me\), in an extreme accelerator, \(\eta\sim1\), produce a synchrotron
  component that reaches the VHE domain. A modest Doppler boosting, \({\cal D}\sim30\), can relax the requirement for
  acceleration efficiency or extend the spectrum to multi-TeV energies. The high cutoff energy of proton synchrotron
  emission comes at a price of very large energy of emitting particles. The energy of particles that reach the
  synchrotron cooling limit scales as \(m^2\), thus proton synchrotron scenarios require a very strong magnetic field of \(B\geq100\rm\,G\) needed to confine the ultra-high energy protons \citep{2000NewA....5..377A}. }

Variations in other underlining assumptions may also change the position of the burnoff limit. For example, if a particle
is on a trajectory closely following a bent magnetic field line then  energy losses of the particle and its peak emitting frequency do
not depend on the strength of the magnetic field but rather on its curvature radius. This emission mechanism is dobbed in
literature as {\it curvature radiation}, and it is feasible that curvature radiation peaks above the synchrotron burnoff
limit.  Acceleration of particles for efficient emission via curvature radiation requires the presence of electric field
directed along the magnetic field. This implies a violation of the ideal MHD conditions at least in certain regions of
the source (in so-called {\it gaps}). Alternatively, one can consider the curvature radiation produced by secondary
electrons, e.g., generated in pulsar magnetosphere by photon magnetic absorption. In this case the initial momentum of
secondary particles is not necessary directed along the magnetic field, so the bulk of particle energy is emitted via
the synchrotron channel.  The ``100~MeV'' limit does not concern the synchrotron radiation produced by secondary
electrons independently on the specific mechanism of their generation. For example, under suitable condition the
products of interactions of primary protons with the surrounding gas or radiation fields can generate a synchrotron
component that extends beyond the burnoff limit.

Models involving the synchrotron emission by secondary leptons represent an extreme realization of scenarios where the
site of particle acceleration and emission are physically separated. Such scenarios do not necessary require violation
of the {ideal} MHD conditions. If a considerable gradient of the magnetic field is present in the source, then particles
accelerated in regions of weak magnetic field can travel up to regions with strong magnetic fields and eventually
produce synchrotron component extending beyond the burnoff limit. Some models invoke conversion between {charged} and
uncharged state of leading particle (e.g., via a cycle of pair production and inverse Compton scattering in the
Klein-Nishina regime), which allows avoiding synchrotron losses during the propagation to a strong B-field region.

Efficiency of acceleration process influences the position of the cutoff of the synchrotron spectrum. In the case of
stochastic particle acceleration, the spectrum of electrons does not stop abruptly but extends beyond $\Ee\ma$, although
with a rather sharp cutoff shape, e.g., exponential or faster. Typically, the maximum of the SED, $\nu F_\nu$, appears
at the energy exceeding $h \nu\ma$. However, for any realistic spectrum of electrons, the shift does not exceed a factor
of a few. If one considers acceleration processes operating in non-ideal MHD configurations, then one may expect high
acceleration efficiencies, \(\eta<1\), and the synchrotron spectrum can extend considerably above the burnoff
limit. Reconnection of magnetic field lines represents a non-ideal MHD process that is believed to play important role
in particle acceleration in different astrophysical sources, including pulsar wind nebulae and jet from active galactic
nuclei. We note that there is a quite broad spectrum of processes that can get activated at magnetic reconnection and
enhance particle acceleration. For example, this includes Fermi I acceleration in converging flows mediated by magnetic
reconnection \cite{2012A&A...542A.125B}, transformation of turbulent cascade with an important feedback on particle
acceleration \cite{2020PhPl...27a2305L}. Among other effects, the acceleration in the current sheet of X-point is of
special interest \citep{2013ApJ...770..147C}. In such regions, the reconnecting magnetic field annihilate leaving guide magnetic field and
unscreened electric field \({\cal E}_0\simeq \beta B_0\), where \(\beta\sim{\cal O}(1)\) is the inflow speed. Although this
configuration reassembles the models with different accelerator and emitter magnetic fields discussed above, we note a
principle differences between these two cases: the reconnection current sheet is not a region with an ideal MHD flow, so
the burnoff limit is not applicable for that configuration.

If the particle confinement time in the current sheet is long enough to accelerate particles beyond \(\Ee\ma\), then these
energetic particles should produce emission that peaks above the burnoff limit once they escape into the regions with
magnetic field \(B_0\). It may seem that scenarios involving particle acceleration in current sheets alleviate completely the constraint imposed by synchrotron cooling. However, in any realistic configuration there are still important further limitations.  Firstly, to generate a synchrotron component that extends in to VHE regime the linear size
of the accelerator should be sufficiently large,
\be
\begin{split}
  \lambda\acc &\geq\quad50~ {\beta}^{-1}\left(\frac{B_{\rm cr}}{B_0\alpha}\right)^{\nicefrac32} \frac{e^2}{\me c^2}\\
  &\geq 2\times10^{17} {\beta}^{-1}\left(\frac{B_0}{1~\rm mG}\right)^{\nicefrac{-3}{2}}\rm\,cm\,,
\end{split}
\ee
where \(B_{\rm cr}\) is the Schwinger critical field strength. Existence of such extended coherent reconnection site is questionable.
Another important factor is the influence of the guide magnetic field, which should be present unless the
reconnecting magnetic fields are perfectly antiparallel. Guide magnetic fields not only lead to synchrotron cooling, but
also enhance escape of high energy particles from the current sheet. These two factors are expected to suppress significantly the efficiency of the synchrotron process in the VHE regime.


Finally, the sharp decrease at the end of the synchrotron spectrum can be altered in a medium with inhomogeneous magnetic fields. For the
case of highly turbulent magnetic fields whose correlation length crossing time, \(\lambda_B/c\), is small compared to
the electron's inverse cyclotron frequency, $1/\omega_c=\me c/ eB_0$, the radiation proceeds in the so-called jitter
regime. In this case, the peak in the radiation spectrum is shifted to higher frequencies by a factor of
\(c/(\omega_c\lambda_B)\) \citep{2013ApJ...774...61K}. Thus, for scenarios with extremely small-scale turbulence, one can
formally expect radiation in the multi-GeV or higher energies. However, the plasma physics community has been quite
pessimistic concerning the condition of $\lambda_B \ll c/\omega_c$, and correspondingly considers the realization of
magnetobremsstrahlung in the jitter regime as a highly unlikely process.  In principle, the presence of large scale
turbulence, \(\lambda_B > c/\omega_c\), can also have an impact on the SED of synchrotron radiation.  In particular, a
broad power-law distribution of the magnetic field strength can result in a rather long power-law tail of radiation
beyond $h\nu\ma$ \citep{2013ApJ...774...61K}.  However, for more realistic magnetic field strength distributions, e.g., a
Gaussian type distribution, the shift of the synchrotron peak is rather mild \citep{2019ApJ...887..181D}.

\section{Toy Model: mathematical formulation} 
\label{sec:toy_math}
\begin{figure}
\plotone{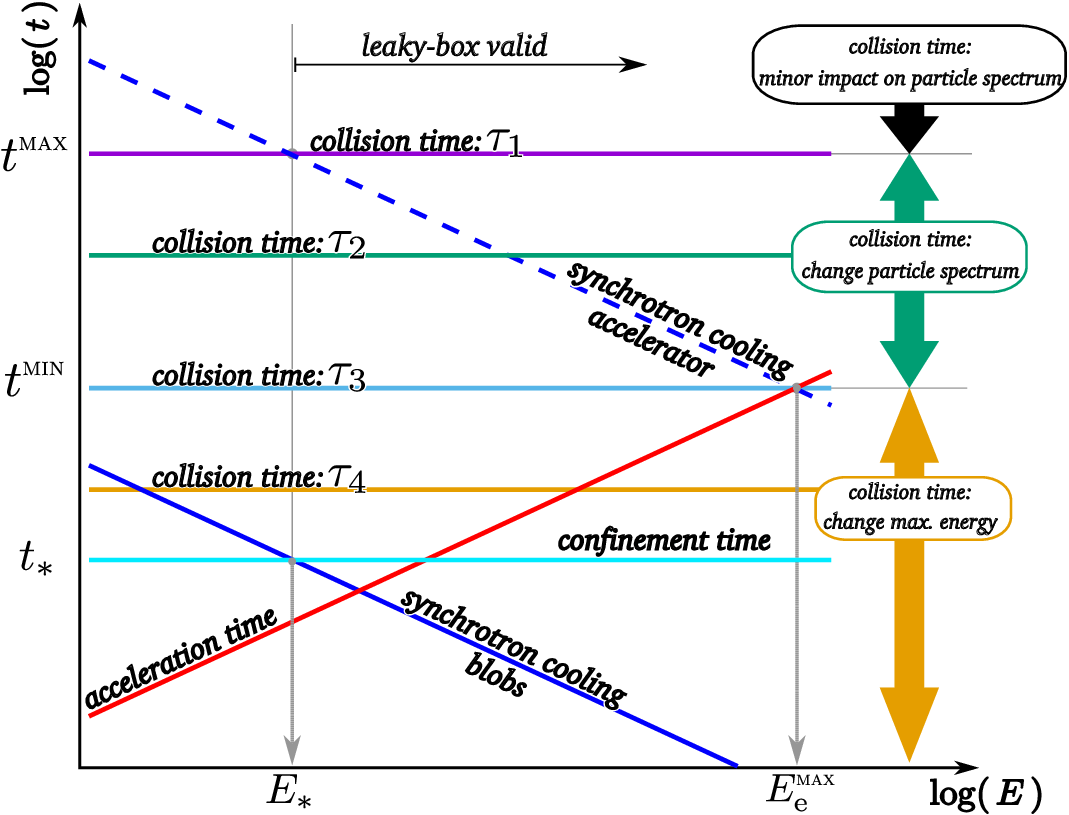}
\caption{\label{fig:time} Characteristic time scales: the acceleration time in the main (acceleration) zone,
\(t\acc\), (red line); the weak field synchrotron cooling time in the main zone, \(t\syn(E,B_0)\), (blue dashed line); the strong field synchrotron cooling time in the blobs, \(t\syn(E,B_*)\), (blue solid line); the confinement time of the electron in the blobs,  \(t_{*}\), (cyan line). Four characteristic collision times are considered: the purple curve for $\tau=t\ma$, the green curve for $\tau=\sqrt{t\mi t\ma}$, the light blue curve for $\tau=t\mi$, and the orange curve for $\tau<t\mi$. Note, the timescales $t\mi$ and $t\ma$ are defined as $t\syn(E\ma,B_{0})$ and $t\syn(E_{*},B_{0})$, respectively.}
\end{figure}

Although the postulation of magnetic field blobs inside the particle accelerator implies that we deal with a multi-zone
scenario, mathematically one can describe the problem within a one-zone model. In this case, the description of the
process is similar to the so-called {\itshape leaky-box approximation} which {has} been successfully applied to Galactic
cosmic rays:
\be \label{eq:leaky}
\frac{\partial n}{\partial t} + \frac{\partial (\dot{E}n)}{\partial E } +\frac{n}{\tau} = q(E)\,.
\ee
{Here \(\tau\) is the particle~--~magnetic blob collision time, and \(\dot{E}\) is the energy loss rate of electrons in the main (acceleration) zone depends on particle energy and magnetic field strength given by
} \be\label{eq:losses}
  \dot{E}(E)=-\frac{E}{t\syn(E,B_0)}=-\frac{\sigmat B_0^2E^2}{6\pi \me^2c^3}\,.
\ee
Within the standard treatment, these terms are responsible for the continuous and catastrophic losses of electrons, respectively. The function 
\(q(E)\)  accounts for a phenomenological description of the acceleration process.  Eq.~\eqref{eq:leaky} is valid for particles with  energy  \(E>E_*\);  the boundary energy is obtained from the solution of the  equation:
\be \label{eq:boundary_energy}
t\syn(E_*,B_*)=t_*\, ,
\ee
where \(B_*\) is the
magnetic field  and  \(t_*\)  is the electron confinement time  in  the blobs. The latter generally depends on the particle energy.  {See Appendix~\ref{ap:leaky_eq} for the justification of Eq.~\eqref{eq:leaky}.}  The
relevant time-scales are shown in Fig.~\ref{fig:time}.  We
assume that in absence of magnetic blobs the injection spectrum contains a cutoff at energy \(E\ma\) which is determined from the condition 
\(t\acc(E\ma)=t\syn(E\ma,B_0)\). We define this timescale $t\mi$. If the collision time is long compared to the acceleration time up to energy \(E\ma\), i.e. 
\be\label{eq:tau_1}
\tau>t\mi\,,
\ee
then  the impact of the magnetic blobs on the maximum energy of accelerated electrons can be ignored. Additionally, it is also helpful to define another timescale $t\ma=t\syn(E_{*},B_{0})$. The spectral shape of electrons in the cutoff region depends on the specific acceleration process. Below,  we assume that \(q\propto E^{-\alpha}{\rm exp}\left[-\left(E/E\ma\right)^\beta\right]\), where for simplicity, but without loss of generality,  we adopt  \(\alpha =2\) and \(\beta=1\).

In the fast cooling regime, {when the injection and losses processes balance one another}, the solution of Eq.~\eqref{eq:leaky} is
\be\label{eq:leaky_particles}
n(E,t)=-\frac1{\dot{E}}\int\limits_{E}^{\infty}\dif{E'}q(E'){\rm exp}\left[-\frac{\tau\eff(E,E')}{\tau}\right]\,,
\ee
where the auxiliary function \(\tau\eff\) is
\be
\tau\eff(E,E')=-\int\limits_{E}^{E'}\frac{\dif{E''}}{\dot{E}(E'')}= \frac{6\pi \me^2c^3}{(\sigmat
  B_0^2)}\left(\frac1{E}-\frac1{E'}\right)\,.
\ee
Two distinct components of synchrotron emission are produced in this scenario. In the 
main zone, the synchrotron radiation produced by electrons with an energy distribution given by Eq.~\eqref{eq:leaky_particles}, takes a spectral shape of the form
\be
L_{\nu,\rm\,0}=\int P_{\nu}(B_0,E) n(E) \dif{E}\,,
\ee
where \(P_\nu\) is the power per unit frequency emitted by an electron with energy \(E\). The electrons that enter the blob, with its much stronger magnetic field, quickly emit their energy at a rate  \(1/\tau\). The time-averaged energy spectrum of their synchrotron radiation produced in the magnetic blobs is  
\be\label{eq:L_nu}
L_{\nu,\rm\,*}=\frac{1}{\tau}\mathlarger{\int}\left[ \int\limits_0^{t_*} \dif{t}P_\nu\Big(B_*,\tilde{E}(t,B_*)\Big) \right]n(E)\dif{E}\,,
\ee
where \(\tilde{E}(t,B)=E/(1+\nicefrac{t}{t\syn(E,B)})\) ({see Appendix~\ref{ap:emission}}).

As demonstrated in Fig.~\ref{fig:time}, for the condition given by  Eq.~\eqref{eq:tau_1}  two different scenarios exist: \(\tau\geq t\ma\) and \(t\mi<\tau<t\ma\). In the first of these regimes, the presence of magnetic field blobs has only a minor impact on the particle spectrum, i.e., \(\tau\gg\tau\eff\). In the second regime the particle spectrum is deformed by interactions with the magnetic blobs, although the maximum energy of the particle spectrum remains unaffected. 

\begin{figure}
\plotone{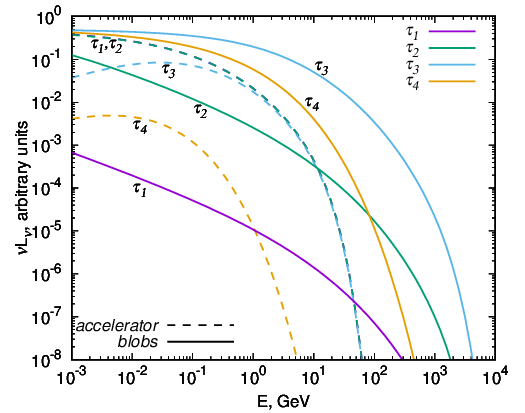}
\caption{\label{fig:SED} Characteristic spectral energy distributions for four example cases considered in Fig.~\ref{fig:time} (the same line colors for the cases considered in Fig.~\ref{fig:time} are used here also). The dashed lines indicate the emission from particles in the accelerator  {magnetic field ($B_{0}=0.1\rm\,G$), and the solid lines indicate the emission from particles in the blobs of size \(R=3\times10^{13}\rm cm\) having magnetic field of $B_{*}=10^2B_{0}$}.
}
\end{figure}

\begin{figure}
\plotone{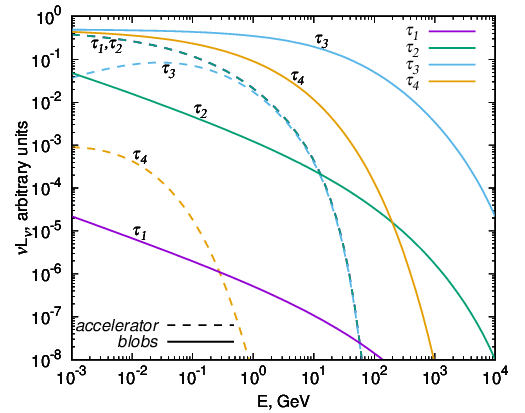}
\caption{\label{fig:SED2}The same as Fig. \ref{fig:SED} but for \(B_*=10^3B_0\).}
\end{figure}

In Figs.~\ref{fig:SED} and \ref{fig:SED2}, we show the corresponding SEDs produced in four different collision timescale scenarios illustrated in Fig.~\ref{fig:time}. In these figures two different strengths of the blob magnetic field, \(B_*=10^2B_0\) and \(B_*=10^3B_0\), are considered. These results show that for particular values of the collision time considered, the synchrotron radiation produced in the blobs indeed extends up to a factor \(B_*/B_0\) beyond the limit $h\nu\ma$. However, for the case of \(\tau>t\ma\), the overall luminosity of this high energy emitted component is significantly less than  that produced in the main zone. Only for the case of small collision timescales,   \(\tau<t\ma\), is the SED dominated by the emission produced in the blobs. The dashed curves in these figures correspond to the emission in the accelerator's magnetic field $B_{0}$ and the solid curves correspond to the emission in the blob's magnetic field $B_{*}$. If particles enter the strong magnetic field region during the acceleration process then the maximum attainable energy can be significantly affected, decreasing the energy of the SED maximum formed in the blobs (the case of \(\tau_4\) in Figs. \ref{fig:SED}~and~\ref{fig:SED2}).

\section{Physical justification of the toy model}
\label{sec:toy_phys}
The mathematical model introduced in Sec.~\ref{sec:toy_math} formally depends on a single parameter, \(\tau\). However,
there are two hidden parameters, \(t_*\) and \(B_*\), which determine the energy range where the leaky box approximation
used is actually valid. Furthermore, the main zone is characterized by two parameters: the size, \(R\) and the magnetic
field, \(B_0\). The question remains whether there exists a physically meaningful combination of these parameters which
would allow the realization of the radiation scenario illustrated in Fig.~\ref{fig:SED}.

For the sake of simplicity, we assume that there exist \(\cal N\) identical small scale large field blobs of size
\(r_*\).  The interaction time of electrons with these blobs lasts a minimum duration of \(t_*\simeq r_*/c\). This
interaction time can be significantly longer if the electron is magnetically confined within the blob.  To avoid this
complication, we focus on sufficiently high electron energies (above $E_{*}$) such that the electron life-time is
shorter than their minimal confinement time in the blobs \(\sim r_*/c\).

For the strong field of the blobs, the geometrical size of the blob dictates the collision time of the electrons with them, 
\be\label{eq:tau}
\tau \simeq \frac{R^3}{{\cal N} \pi r_*^2 c}\,,
\ee
where we used \(R^3\) for the volume of the main zone.
On the other hand, the magnetic field strength of the blob determines the electron energy above which the leaky-box 
treatment remains valid:
\be\label{eq:E_*}
E_*=\frac{6\pi \me^2c^4}{\sigmat r_*B_*^2}\,.
\ee
The scenario introduced in Sec.~\ref{sec:toy_math} is realized when the following three conditions are satisfied:
(i) \(t\mi<\tau<t\ma\), (ii) \(E_*\ll E\ma\), and (iii)  \(B_*\gg B_0\). The former condition determines the range for
\(\cal N\), and can be fulfilled without interference with the two other conditions. We therefore need to check just one condition, \(E_*\ll E\ma\), in the limit \(B_*\gg B_0\):
\be\label{eq:r_*}
r_*\gg \frac{B_0^2}{B_*^2}\frac{E\ma\eta}{eB_0}\,.
\ee
From the application of the maximum energy condition for a source (``Hillas criterion'')  \({E\ma}/({eB_0})<R\),  it follows that there should exist a range of \(r_*\) such that \(E_*\ll E\ma\) and \(r_*\ll R\). However, this requires a large ratio of magnetic fields: \(B_*\gg B_0\).

Let us  estimate the fraction of the magnetic energy in the blobs
required for strong boosting of the radiation energy. It is determined by the condition \(\tau\ll t\ma\):
\be
\frac{\delta w_{\rm B}}{w_{\rm B}} = \frac{4\pi{\cal N}}3 \frac{r_*^3B_*^2}{R^3B_0^2} = \frac43\frac{t\ma}{\tau}\geq1\,.
\ee
This implies that significant transformation of the synchrotron spectrum beyond $h\nu\ma$ is possible when the magnetic energy in the small blobs dominates over the energy in the main zone magnetic field. Note that  this condition can be realized only in weakly magnetized environments, where the energy contained in the particles can provide an ample source for the formation of strong magnetic fluctuations (blobs). {If the particle internal energy is transferred to the magnetic form in blobs, then the total pressure in the blobs should not exceed the external pressure, preventing  such regions of strong magnetic field from instantly expanding.

We note that such conditions with weakly magnetized plasma are expected in several important gamma-ray sources. For example, while pulsar winds are believed to be weakly magnetized close to the termination shock, the overall conditions in pulsar wind nebulae favor equipartition of gas and magnetic pressure. Thus, initially weak magnetic field gets significantly amplified in the nebula. Another example could be gamma-ray burst afterglow emission. According to the standard framework, the afterglow emission is generated by particles that are accelerated at the GRB forward shock that propagates through the circumburst medium. If the circumburst medium  magnetic field is \(\sim10\rm\upmu G\), then the initial plasma magnetization is extremely small, say \(10^{-10}\). The amplification of such a weak magnetic field to some ``equipartition'' strength may create conditions favorable for the formation of compact magnetic blobs.}

\section{Magnetic mirroring}\label{sec:mir}

{Another important aspect is the ability of particles to enter the regions of strong magnetic field. If the change of magnetic field is small, then particles moves in a way that the adiabatic invariant \(p_\perp^2/B\) is approximately conserved. In this case the conservation of energy prevents particle penetration into a region with a significantly stronger magnetic field. To take account of this effect, which is known as magnetic mirroring, and allow particle penetration from a region with magnetic field \(B_0\), the following condition should be fulfilled:}
\be\label{eq:mirror}
T\dif[t]{B}\gtrsim B_0\,,
\ee
{where \(T\sim 2\pi E/eB_0c\) and \(\dif[t]{B}\sim cB_*/r_*\) are the particle giro-rotation period and effective rate of the magnetic field change. Solving Eq.~\ref{eq:mirror} for particle energy, we obtain the minimum energy of particles that can penetrate into the blobs:}
\be\label{eq:mirror}
E\mi = \frac1{2\pi}\frac{B_0}{B_*} r_*eB_0\,.
\ee
{The realization of the discussed scenario requires that \(E\mi\ll E\ma\), which can be written as a condition on the blob size:}
\be\label{eq:r_mirror}
r_*\ll \frac{2\pi B_*}{B_0}\frac{E\ma}{eB_0}\,.
\ee
{For \(B_*\gg B_0\), Eqs.~\eqref{eq:r_*}~and~\eqref{eq:r_mirror} can be simultaneously fulfilled. }

Equations \eqref{eq:r_mirror} and \eqref{eq:r_*_range} show that for the relevant range of magnetic blob sizes, the magnetic mirroring effect should not prevent penetration of particle into the regions of strong magnetic field. However, even if a particle penetrates to the blob, it doesn't necessarily mean that it could interact with the strongest field. For any realistic configuration of the field, there should exist a gradual increase from the main-zone field strength, \(B_0\), to the blob field, \(B_*\). Motion of a particle in this boundary layer might be very complicated. Although the adiabatic invariant should not be conserved at a strong gradient of magnetic field, we may use it to estimate how deep a particle can penetrate into the blob. We assume that the strongest magnetic field strength, \(B(E)\), a particle with energy \(E\) can reach, is determined by a relation similar to Eq.~\eqref{eq:mirror}:
\be\label{eq:mirror2}
\frac{2\pi E}{e B(E)c} \frac{B_*c}{r_*}\simeq B(E)\,.
\ee
We note that this condition corresponds to the strongest magnetic mirroring. The level of the real effect might be significantly smaller. For example, observations of gamma-ray emission from the Sun \citep{2018PhRvL.121m1103L} suggest that magnetic mirroring is not as efficient for reflecting high-energy particles as it was naively expected \citep{1991ApJ...382..652S}.

Solving Eq.~\eqref{eq:mirror2} for \(B(E)\), we obtain:
\be\label{eq:B_on_E}
B(E) \simeq \sqrt{\frac{2\pi E B_*}{e r_*}}\,.
\ee
In addition one should take into account that the minimum and maximum magnetic field strength are \(B_0\) and \(B_*\), respectively.

\begin{figure}
\plotone{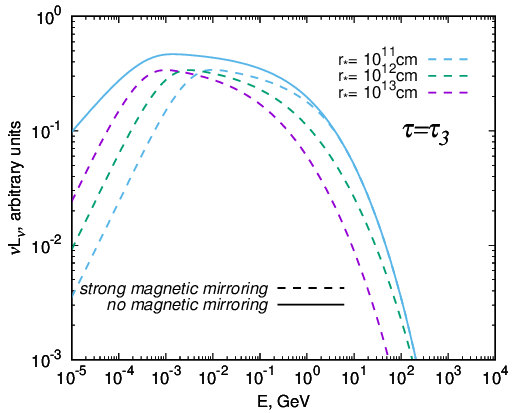}
\caption{\label{fig:mir} Characteristic spectral energy distributions from magnetized blobs showing the impact of strong magnetic mirroring for {\(\tau=\tau_3\) and} three different blob sizes: \(r_*=10^{13}\rm\,cm\), \(r_*=10^{12}\rm\,cm\), and \(r_*=10^{11}\rm\,cm\). The rest of the parameters are the same as in Fig. \ref{fig:SED}.}
\end{figure}

\begin{figure}
\plotone{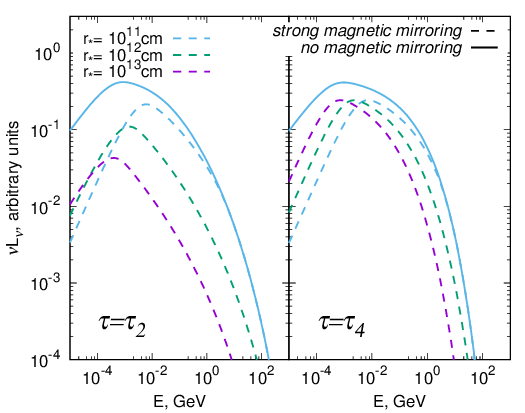}
\caption{\label{fig:mir2} The same as Fig. \ref{fig:mir} but for \(\tau=\tau_2\) (left panel) and \(\tau=\tau_4\) (right panel).}
\end{figure}

To estimate the strongest possible impact of the magnetic {mirroring}, we compute the spectrum from the blob using the energy dependent magnetic field strength, Eq.~\eqref{eq:B_on_E}:
\be
L_{\nu,\rm\,*}^{\rm mir}=\frac{1}{\tau}\mathlarger{\int}\left[ \int\limits_0^{t_*} \dif{t}P_\nu\Big(B(E),\tilde{E}\big(t,B(E)\big)\Big) \right]n(E)\dif{E}
\ee
(compare to Eq.~\eqref{eq:L_nu}). In Figs.~\ref{fig:mir} and ~\ref{fig:mir2} we show the corresponding SEDs for three different blob sizes: \(r_*=10^{13}\rm\,cm\), \(r_*=10^{12}\rm\,cm\), and \(r_*=10^{11}\rm\,cm\). As can be seen, in the case of strong magnetic mirroring, the spectra can become significantly suppressed at low energies. This is caused by significant reflection of low-energy electrons. Higher energy particles may have difficulties in reaching the strongest magnetic field, which results in spectrum suppression for large blobs, \(r_*\geq10^{12}\rm \,cm\). 

\section{Discussion}

The synchrotron radiation of relativistic electrons is a unique process that proceeds with high efficiency in a wide variety of astrophysical environments, and spans a broad energy range from radio up to gamma-rays. The typical short time-scales characterizing the synchrotron cooling of electrons make this process the dominant radiation channel not only at radio wavelengths but also in the relatively low (Sub-GeV) energy gamma-ray band.  This is the case, for example, of gamma-ray bursts \citep{2018ApJ...863..138A} or the so-called Crab flares \citep{2014RPPh...77f6901B}.  The characteristic spectral and temporal features of this mechanism seem very attractive for interpretation of the fast and energetic phenomena in the very high energy gamma-ray band as well.  One may recall, for example, the ultrafast VHE gamma-ray flares of blazars \citep{2007ApJ...664L..71A,2014Sci...346.1080A} or extension of the gamma-ray spectra of late afterglows of GRBs \citep{2019Natur.575..464A}. However, the self-regulated cutoff in the synchrotron spectrum at $\leq$100 MeV has been thought to limit the applicability of the process for energies well above 100~MeV (in the frame of the source). 

Synchrotron radiation can be significantly boosted in sources where the emission region moves relativistically, $h\nu\ma \simeq 70 \ {\cal D} \ \rm MeV$. The Doppler boosting of the maximum of synchrotron radiation can be naturally realized in relativistic outflows like the jets of active galactic nuclei (AGN) and gamma-ray bursts (GRBs). Typically, in the AGN and late GRB afterglows, the jet's Doppler factors do not exceed 100. Thus the Doppler boosting alone cannot shift the synchrotron radiation to very-high energies, especially given that in realistic conditions, e.g. with the inclusion of the presence of dissipative factors like adiabatic losses or plasma instabilities, the $\eta$-parameter may significantly exceed 1. The consideration of physically realistic values for $\eta$, therefore, exacerbates further the problem of accounting for emission at the $h\nu\ma$ scale as being synchrotron in origin.

In this paper, we propose a scenario for the formation of synchrotron radiation in compact magnetic blobs located inside the particle accelerator. The blobs do not participate in the acceleration process but serve as passive targets where the highest energy electrons accelerated in the main zone, are instantly cooled via synchrotron losses, whose spectrum is effectively extended up to the VHE band.  We have developed a simple mathematical formalism based on a ``leaky box'' approach, for calculations of synchrotron radiation emerging from such systems. The results of numerical calculations confirm the qualitative estimates. They demonstrate that the synchrotron radiation of such systems can be effectively released in the very-high-energy regime provided that the magnetic field in the blobs exceeds two or three orders of magnitude the magnetic field in the accelerator zone. Although the compact magnetic condensations occupy a rather small volume inside the accelerator, the total magnetic energy in these blobs should significantly exceed the electromagnetic energy contained in the acceleration zone.  Remarkably, the MHD studies of magnetic 
amplification by turbulent dynamo \citep{2009ApJ...692L..40Z} in relativistic  outflows may result in such systems where the magnetic energy is predominantly contained in compact clumps. Moreover, this amplification of small scale fields may be a general effect \citep{1968JETP...26.1031K}.

As shown in Sec.~\ref{sec:toy_phys}, the hierarchy of the characteristic timescales determining the feasibility of the scenario is such that the extension of the blob synchrotron component to the VHE regime is possible. Let us here outline some criteria for that regime. The  toy model from Sec.~\ref{sec:toy_phys} contains six parameters: \(R\), \(B_0\), \(\eta\), \(r_*\), \(\cal N\), and \(B_*\). It is convenient to replace parameter \(\cal N\) with {\it filling factor}, i.e., \({\cal F} =  \nicefrac{4\pi}{3}\, {\cal N}(r_*/R)^3=\nicefrac43\,r_*/(\tau c)\), as it allows us to eliminate the dependences on the size of the system, \(R\). Thus, the radiation regime depends on only five parameters. 

The extension of the synchrotron component allows one to define the strength of the blob magnetic field: \(B_*\sim 10^3 B_0\eta\). This component has a significant luminosity if \(\tau\simeq t\mi\), which translates into condition \({\cal F}\sim \nicefrac43\, r_*/(t\mi c)\). These two conditions further reduce the number of free parameters. Finally, two conditions, \({\cal F}\ll 1\) and Eq.~(\ref{eq:r_*}), define the allowed range for \(r_*\):
\be\label{eq:r_*_range}
\frac34 \frac{\eta \Ee\ma}{eB_0}\gg r_* \gg  10^{-6}\frac{\Ee\ma}{eB_0\eta}\,,
\ee
where \(\Ee\ma\) depends on \(B_0\) and \(\eta\):
\be
\Ee\ma=\sqrt{\frac{6\pi e\me^2c^4}{\sigmat \eta B_0}}=2\left(\frac{\eta B_0}{1\rm\,mG}\right)^{\nicefrac{-1}{2}}\rm\,PeV\,.
\ee
Normalizing the magnetic field to \(1\rm mG\), one obtains
\be\label{eq:r_*_range2}
\frac34 \eta^{\nicefrac12}\left(\frac{B_0}{1\rm\,mG}\right)^{\nicefrac{-3}{2}}\gg \frac{r_*}{ 6\times10^{15}{\rm \,cm}} \gg   10^{-6}\eta^{\nicefrac{-3}{2}}\left(\frac{B_0}{1\rm\,mG}\right)^{\nicefrac{-3}{2}}\,.
\ee
This multi-decade range shows that the requirements for the discussed scenario can be relatively easily met in astrophysical sources. {We note here that Eq.~\ref{eq:r_mirror} provides a weaker constraint on the blob size compared to Eq.~\eqref{eq:r_*_range}. Furthermore, even for the extreme case of effective magnetic mirroring, this scenario can still be realized provided that the blobs have a sufficiently small size:
\be\label{eq:r_*_range2}
10^{-2}\eta^{\nicefrac{-3}{2}}\left(\frac{B_0}{1\rm\,mG}\right)^{\nicefrac{-3}{2}}\gtrsim \frac{r_*}{ 6\times10^{15}{\rm \,cm}} \gg   10^{-6}\eta^{\nicefrac{-3}{2}}\left(\frac{B_0}{1\rm\,mG}\right)^{\nicefrac{-3}{2}}\,.
\ee
}

\acknowledgments
We thank anonymous referee for useful comments that helped us to improve the manuscript.
DK acknowledges support by JSPS KAKENHI Grant Numbers 18H03722, 18H05463, and 20H00153.

\appendix

\section{Leaky-box approximation} \label{ap:leaky_eq}
Magnetic field
fluctuations play a key role in both sink and source terms in the standard injection-cooling equation:
\be\label{eq:inj_cool}
\frac{\partial n}{\partial t} + \frac{\partial (\dot{E}n)}{\partial E } +\frac{n}{\tau} - \frac{n(\tilde{E})}{\tau} = q(E)\,,
\ee
where \(q(E)\) is the injection term, \(n=\dif[E]{N}\) is energy-distribution function, \(\dot{E}\) is
energy losses in the main zone magnetic field, \(\tau\) is the particle~--~fluctuation collision time,
\(\tilde{E}(E)\) is the particle energy that cools down to \(E\) while interacting with the
fluctuation. While the particle is interacting with the fluctuation, it loses energy via synchrotron emission in the strong magnetic field of strength \(B_*\). If the {confinement} time is \(t_*\), {then the initial and final particle energies (\(\tilde{E}\) and \(E\), respectively)are related as}
\be\label{eq:e_tilde}
\frac{E}{E_*} = \frac{\frac{\tilde{E}}{E_*}}{1+\frac{\tilde{E}}{E_*}}\,,
\ee
{where \(t\syn(E,B_*)(\propto E^{-1}B_*^{-2})\) is the synchrotron cooling time in the
  magnetic field of strength \(B_*\) and \(E_*\) is the solution of equation \(t_*=t\syn(E_*,B_*)\).}
{According to Eq.~\eqref{eq:e_tilde}, for the energy range \(E>E_*\), there are no particles that cool down to \(E\) and Eq.~\eqref{eq:inj_cool} takes a simple form}
\be
\frac{\partial n}{\partial t} + \frac{\partial (\dot{E}n)}{\partial E } +\frac{n}{\tau} = q(E)\,,
\ee
which is identical to the equation for transport of cosmic ray under the leaky-box approximation. It allows a well-known analytical solution \citep{1959SvA.....3...22S}: 
\be
n(E,t)=-\frac1{\dot{E}}\int\limits_{E}^{E\eff}\dif{E'}q(E'){\rm exp}\left[-\frac{\tau\eff(E,E')}{\tau}\right]\,.
\ee
Here the auxiliary parameters, \(E\eff\) and \(\tau\eff\), are  
\be
t=-\int\limits_E^{E\eff}\frac{\dif{E'}}{\dot{E}(E')}
\ee
and
\be
\tau\eff(E,E')=-\int\limits_{E}^{E'}\frac{\dif{E''}}{\dot{E}(E'')}\,.
\ee
Since the dominant cooling mechanism is synchrotron losses the above integral can be derived analytically. 

\section{Approximated expression for the synchrotron spectrum}\label{ap:emission}
\setcounter{equation}{0}
The total emission of a particle that enters a high magnetic field region needs to be computed accounting for the change of the particle energy.  To be consistent with the standard textbooks, we operate with the total emitted energy per frequency:
\be\label{eq:total_emission}
E_\nu=\int\limits_0^{t_*} P_\nu\big(E(t,B)\big)\dif{t}=\frac{\sqrt{3}e^3B}{\me c^2}\int\limits_0^{t_*} F_\nu\big(x(t,B)\big)\dif{t}\,,
\ee
where
\be
F(x)=x\int\limits_x^\infty K_{\nicefrac53}(\xi)\dif{\xi}\,
\ee
and
\be
x(t,B)=\frac{4\pi\nu \me^3c^5}{3E_0^2eB}u^2=\nu_0u^2(t,B)
\ee
(here \(u(t,B)=\nicefrac{E_0}{E(t,B)}\) and the above equation defines \(\nu_0\)).
In Eq.~\eqref{eq:total_emission} the particle energy, \(E(t)\), depends on the initial energy, \(E_0\), as
\be\label{eq:cooling}
E(t,B)=\frac{E_0}{1+t/t\syn(E_0,B)}
\ee
and
\be
\dif{t}=t\syn(E_0,B)\dif{u}\,.
\ee
 Although, formally, Eq.~\eqref{eq:total_emission} is a four-parameter function, it is clear that
\be
E_\nu\big(E_0,B,t_*\big)=E_\nu\big(E_0,B,\infty\big) - E_\nu\big(E(t_*,B),B,\infty\big)\,,
\ee
which allows one to reduce the dependence on \(t_*\), and in what follows we consider Eq.~\eqref{eq:total_emission} in the limit \(t_*\rightarrow\infty\).

\begin{figure}
\plotone{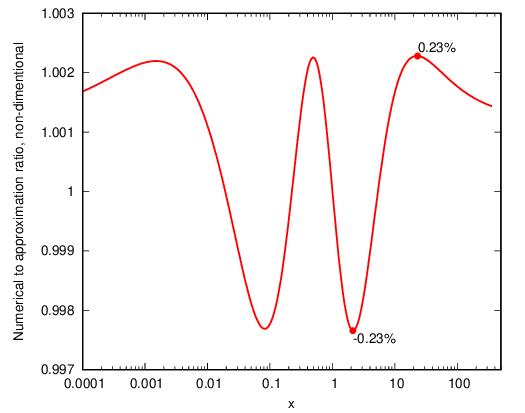}
\caption{\label{fig:approx} Ratio of the exact and approximate function that describe the synchrotron spectrum accounting for change of the particle energy.}
\end{figure}

Combining the above equations, one obtains
\be
\begin{split}
  E_\nu&=\frac{\sqrt{3}e^3Bt\syn(E_0,B)}{\me c^2}\int\limits_1^{\infty}\dif{u}\nu_0 u^2\int\limits_{\nu_0 u^2}^{\infty}\dif{\xi}K_{\nicefrac53}(\xi)\\
  &=\frac{\sqrt{3}e^3B_*t\syn(E_0,B)}{ \me c^2}G(\nu_0)\,.
\end{split}
\ee
Changing the integral order allows one to derive one integral in the auxiliary function analytically:
\be\label{eq:G_fun}
G(\nu_0)=\frac{\nu_0}3\int\limits_{\nu_0}^{\infty}\dif{\xi}K_{\nicefrac53}(\xi)\left(\left(\frac{\xi}{\nu_0}\right)^{\nicefrac32}-1\right)\,.
\ee
Using the approach introduced in \citet{2014ApJ...783..100K}, we approximate Eq.~\eqref{eq:G_fun} with a simple analytic function:
\be\label{eq:G_approx}
G(x)\simeq \tilde{G}(x)=\frac{1.04{\rm e}^{-x}}{\sqrt{x}}\left(\frac{1+0.464x}{1+0.771x}\right)g(x)\,,
\ee
where
\be
g(x)= \left[1+\frac{a x^\alpha}{1+bx^\beta}\right]^{-1}
\ee
for \(a=0.186\), \(\alpha=0.57\), \(b=1.75\), and \(\beta=1.49\).
This approximation has exact asymptotical behavior for \(x\rightarrow 0\) and \(\infty\). In the entire range it provides an accuracy of \(\simeq0.2\%\) as shown in Fig.~\ref{fig:approx}.


\end{document}